\documentclass[aps,preprint,nofootinbib,floatfix]{revtex4}%
\usepackage{amssymb}
\usepackage{amsmath}
\usepackage{epsfig}
\usepackage{amsfonts}
\usepackage{graphicx}%
\begin{document}
\title{Linear Relations and their Breakdown in High Energy Massive String Scatterings
in Compact Spaces}
\author{Jen-Chi Lee}
\altaffiliation{Email: jcclee@cc.nctu.edu.tw}

\altaffiliation{(on leave of absence from NCTU)}

\affiliation{Institute of Physics, University of Tokyo, Komaba, Meguro-ku, Tokyo 153-8902, Japan}
\affiliation{Department of Electrophysics, National Chiao-Tung University, Hsinchu, Taiwan, R.O.C.}
\author{Yi Yang\thanks{Email: yyang@phys.cts.nthu.edu.tw (Physics Division, National
Center for Theoretical Sciences, Hsinchu, Taiwan, R.O.C.)}}
\altaffiliation{Email: yyang@phys.cts.nthu.edu.tw}

\altaffiliation{(on leave of absence from NCTS)}

\affiliation{Institute of Physics, University of Tokyo, Komaba, Meguro-ku, Tokyo 153-8902, Japan}
\affiliation{Physics Division, National Center for Theoretical Sciences, Hsinchu, Taiwan, R.O.C.}
\date{\today }

\begin{abstract}
We calculate high energy massive scattering amplitudes of closed bosonic
string compactified on the torus. For each fixed mass level with given
quantized and winding momenta $\left(  \frac{m}{R},\frac{1}{2}nR\right)  $, we
obtain infinite linear relations among high energy scattering amplitudes of
different string states. For some kinematic regimes, we discover that linear
relations with $N_{R}=N_{L}$ break down and, simultaneously, the amplitudes
enhance to power-law behavior instead of the usual expoential fall-off
behavior at high energies. It is the space-time T-duality symmetry that plays
a role here. This result is consistent with the coexistence of the linear
relations and the softer exponential fall-off behavior of high energy string
scattering amplitudes as we pointed out prevously. It is also reminiscent of
our previous work on the power-law behavior of high energy string/domain-wall scatterings.

\end{abstract}
\maketitle
\tableofcontents
%

\setcounter{equation}{0}
\renewcommand{\theequation}{\arabic{section}.\arabic{equation}}%

\section{Introduction}

It is well known that there are two fundamental characteristics of high energy
string scattering amplitudes, which make them very different from field theory
scatterings. These are the softer exponential fall-off behavior (in contrast
to the power-law behavior of field theory scatterings) and the existence of
infinite Regge-pole structure in the form factor of the high energy string
scattering amplitudes.

Recently high-energy, fixed angle behavior of string scattering amplitudes
\cite{GM, Gross, GrossManes} was intensively reinvestigated for massive string
states at arbitrary mass levels \cite{ChanLee1,ChanLee2,
CHL,CHLTY,PRL,paperB,susy,Closed,HL}. An infinite number of linear relations,
or stringy symmetries, among string scattering amplitudes of different string
states were obtained. An important new ingredient of these calculations is the
zero-norm states (ZNS) \cite{ZNS1,ZNS3,ZNS2} in the old covariant first
quantized (OCFQ) string spectrum. The existence of these infinite linear
relations constitutes the \textit{third} fundamental characteristics of high
energy string scatterings, which is not shared by the usual point-particle
field theory scatterings.

These linear relations persist for string scattered from generic D$p$-brane
\cite{Dscatt} except D-instanton and Domain-wall. For the scattering of
D-instanton, the form factor exhibits the well-known power-law behavior
without Regge-pole structure, and thus resembles a field theory amplitude. For
the special case of Domain-wall (D24-brane) scattering, it was discovered
\cite{Wall} recently that, in contrast to the common wisdom of exponential
fall off behavior \cite{Klebanov,Myers}, its form factor behaves as\textit{
power-law} with Regge-pole structure. This discovery makes Domain-wall
scatterings a hybrid of string and field theory scatterings. Moreover, it was
shown \cite{Wall} that the linear relations break down for the Domain-wall
scattering due to this unusual power-law behavior. This result gives a strong
evidence that the existence of the infinite linear relations, or stringy
symmetries, of high-energy string scattering amplitudes is responsible for the
softer (exponential fall-off) high-energy string scatterings than the
(power-law) field theory scatterings.

To further convince ourselves with the coexistence of the infinite linear
relations and the softer exponential fall-off behavior of string scatterings
at high energies, it is important to find more examples of high energy string
scatterings, which show the unusual power-law behavior and, simultaneously,
give the breakdown of the infinite linear relations. With this in mind, in
this paper we calculate high energy massive scattering amplitudes of closed
bosonic string with some coordinates compactified on the torus
\cite{Mende,Khuri}. For each fixed mass level with given quantized and winding
momenta $\left(  \frac{m}{R},\frac{1}{2}nR\right)  $, we obtain infinite
linear relations among high energy scattering amplitudes of different string
states. This result is reminiscent of the existence of an infinite number of
massive soliton ZNS in the compactified string constructed in \cite{Lee}. We
then discover that, for some kinematic regime, so called Mende regime (MR),
infinite linear relations with $N_{R}=N_{L}$ break down and, simultaneously,
the amplitudes enhance to power-law behavior instead of the usual exponential
fall-off behavior at high energies. It is the space-time T-duality symmetry
that plays a role here.

The power-law behavior of high energy string scatterings in a compact space
was first suggested by Mende \cite{Mende}. Here we give an explicit
calculation of the conjecture. Moreover, in addition to the high energy
string/domain-wall scatterings mentioned above \cite{Wall}, our result in this
paper gives another evidence to support the coexistence of the infinite linear
relations and the softer exponential fall-off behavior of high energy string
scattering amplitudes as we pointed out previously \cite{Wall, Decay}. The
result also suggests that the infinite linear relations are closely related to
the full 26D space-time symmetry of closed bosonic string theory. This paper
is organized as following. In section II, we set up the kinematic for the
compactified string and calculate the four-tachyon $(N_{R}=N_{L}=0)$
scattering amplitudes with arbitrary winding. In section III, we derive the
infinite linear relations among high energy scattering amplitudes of different
string states with given $\left(  \frac{m}{R},\frac{1}{2}nR\right)  $ for each
fixed mass level. We then discuss the power-law behavior of the amplitudes and
breakdown of the infinite linear relations in the Mende regime. A brief
conclusion is given in section IV.%

\setcounter{equation}{0}
\renewcommand{\theequation}{\arabic{section}.\arabic{equation}}%

\section{String compactified on torus}

\subsection{Winding string and kinematic setup}

We consider 26D closed bosonic string with one coordinate compactified on
$S^{1}$ with radius $R$. As we will see later, it is straightforward to
generalize our calculation to more compactified coordinates. The closed string
boundary condition for the compactified coordinate is (we use the notation in
\cite{GSW})%

\begin{equation}
X^{25}(\sigma+2\pi,\tau)=X^{25}(\sigma,\tau)+2\pi Rn,
\end{equation}
where $n$ is the winding number. The momentum in the $X^{25}$ direction is
then quantized to be%
\begin{equation}
K=\frac{m}{R},
\end{equation}
where $m$ is an integer. The mode expansion of the compactified coordinate is%
\begin{equation}
X^{25}\left(  \sigma,\tau\right)  =X_{R}^{25}\left(  \sigma-\tau\right)
+X_{L}^{25}\left(  \sigma+\tau\right)  ,
\end{equation}
where%
\begin{align}
X_{R}^{25}\left(  \sigma-\tau\right)   &  =\frac{1}{2}x+K_{R}\left(
\sigma-\tau\right)  +i\sum_{k=0}\frac{1}{k}\alpha_{k}^{25}\text{
}e^{-2ik\left(  \sigma-\tau\right)  },\\
X_{L}^{25}\left(  \sigma+\tau\right)   &  =\frac{1}{2}x+K_{L}\left(
\sigma+\tau\right)  +i\sum_{k=0}\frac{1}{k}\tilde{\alpha}_{k}^{25}\text{
}e^{-2ik\left(  \sigma+\tau\right)  }.
\end{align}
The left and right momenta are defined to be%
\begin{equation}
K_{L,R}=K\pm L=\frac{m}{R}\pm\dfrac{1}{2}nR\Rightarrow K=\dfrac{1}{2}\left(
K_{L}+K_{R}\right)  ,
\end{equation}
and the mass spectrum can be calculated to be%
\begin{equation}
\left\{
\begin{array}
[c]{c}%
M^{2}=\left(  \dfrac{m^{2}}{R^{2}}+\dfrac{1}{4}n^{2}R^{2}\right)  +N_{R}%
+N_{L}-2\equiv K_{L}^{2}+M_{L}^{2}\equiv K_{R}^{2}+M_{R}^{2}\\
N_{R}-N_{L}=mn
\end{array}
\right.  , \label{mass}%
\end{equation}
where $N_{R}$ and $N_{L}$ are the number operators for the right and left
movers, which include the counting of the compactified coordinate. We have
also introduced the left and the right level masses as%
\begin{equation}
M_{L,R}^{2}\equiv2\left(  N_{L,R}-1\right)  . \label{level mass}%
\end{equation}
Note that for the compactified closed string $N_{R}$ and $N_{L}$ are
correlated through the winding modes.

In the center of momentum frame, the kinematic can be set up to be%

\begin{align}
k_{1L,R}  &  =\left(  +\sqrt{p^{2}+M_{1}^{2}},-p,0,-K_{1L,R}\right)  ,\\
k_{2L,R}  &  =\left(  +\sqrt{p^{2}+M_{2}^{2}},+p,0,+K_{2L,R}\right)  ,\\
k_{3L,R}  &  =\left(  -\sqrt{q^{2}+M_{3}^{2}},-q\cos\phi,-q\sin\phi
,-K_{3L,R}\right)  ,\\
k_{4L,R}  &  =\left(  -\sqrt{q^{2}+M_{4}^{2}},+q\cos\phi,+q\sin\phi
,+K_{4L,R}\right)
\end{align}
where $p\equiv\left\vert \mathrm{\vec{p}}\right\vert $ and $q\equiv\left\vert
\mathrm{\vec{q}}\right\vert $ and%
\begin{align}
k_{i}  &  \equiv\dfrac{1}{2}\left(  k_{iR}+k_{iL}\right)  ,\\
k_{i}^{2}  &  =K_{i}^{2}-M_{i}^{2},\\
k_{iL,R}^{2}  &  =K_{iL,R}^{2}-M_{i}^{2}\equiv-M_{iL,R}^{2}.
\end{align}
With this setup, the center of mass energy $E$ is%
\begin{equation}
E=\dfrac{1}{2}\left(  \sqrt{p^{2}+M_{1}^{2}}+\sqrt{p^{2}+M_{2}^{2}}\right)
=\dfrac{1}{2}\left(  \sqrt{q^{2}+M_{3}^{2}}+\sqrt{q^{2}+M_{4}^{2}}\right)  .
\label{COM}%
\end{equation}
The conservation of momentum on the compactified direction gives%
\begin{equation}
m_{1}-m_{2}+m_{3}-m_{4}=0, \label{kk}%
\end{equation}
and T-duality symmetry implies conservation of winding number%
\begin{equation}
n_{1}-n_{2}+n_{3}-n_{4}=0. \label{wind}%
\end{equation}
One can easily calculate the following kinematic relations%
\begin{align}
-k_{1L,R}\cdot k_{2L,R}  &  =\sqrt{p^{2}+M_{1}^{2}}\cdot\sqrt{p^{2}+M_{2}^{2}%
}+p^{2}+\vec{K}_{1L,R}\cdot\vec{K}_{2L,R}\label{k1*k2}\\
&  =\dfrac{1}{2}\left(  s_{L,R}+k_{1L,R}^{2}+k_{2L,R}^{2}\right)  =\dfrac
{1}{2}s_{L,R}-\frac{1}{2}\left(  M_{1L,R}^{2}+M_{2L,R}^{2}\right)  ,
\end{align}%
\begin{align}
-k_{2L,R}\cdot k_{3L,R}  &  =-\sqrt{p^{2}+M_{2}^{2}}\cdot\sqrt{q^{2}+M_{3}%
^{2}}+pq\cos\phi+\vec{K}_{2L,R}\cdot\vec{K}_{3L,R}\\
&  =\dfrac{1}{2}\left(  t_{L,R}+k_{2L,R}^{2}+k_{3L,R}^{2}\right)  =\dfrac
{1}{2}t_{L,R}-\frac{1}{2}\left(  M_{2L,R}^{2}+M_{3L,R}^{2}\right)  ,
\end{align}%
\begin{align}
-k_{1L,R}\cdot k_{3L,R}  &  =-\sqrt{p^{2}+M_{1}^{2}}\cdot\sqrt{q^{2}+M_{3}%
^{2}}-pq\cos\phi-\vec{K}_{1L,R}\cdot\vec{K}_{3L,R}\\
&  =\dfrac{1}{2}\left(  u_{L,R}+k_{1L,R}^{2}+k_{3L,R}^{2}\right)  =\dfrac
{1}{2}u_{L,R}-\frac{1}{2}\left(  M_{1L,R}^{2}+M_{3L,R}^{2}\right)
\end{align}
where the left and the right Mandelstam variables are defined to be%
\begin{align}
s_{L,R}  &  \equiv-(k_{1L,R}+k_{2L,R})^{2},\\
t_{L,R}  &  \equiv-(k_{2L,R}+k_{3L,R})^{2},\\
u_{L,R}  &  \equiv-(k_{1L,R}+k_{3L,R})^{2},
\end{align}
with%
\begin{equation}
s_{L,R}+t_{L,R}+u_{L,R}=\sum_{i}M_{iL,R}^{2}. \label{sum}%
\end{equation}

\subsection{Four-tachyon scatterings with $N_{R}=N_{L}=0$}

We are now ready to calculate the string scattering amplitudes. Let's first
calculate the case with $N_{R}+N_{L}=0$ (or $N_{R}=N_{L}=0$),%
\begin{align}
&  A_{\text{closed}}^{(N_{R}+N_{L}=0)}\left(  s,t,u\right) \nonumber\\
&  =\int d^{2}z\exp\left\{  k_{1L}\cdot k_{2L}\ln z+k_{1R}\cdot k_{2R}\ln
\bar{z}+k_{2L}\cdot k_{3L}\ln\left(  1-z\right)  +k_{2R}\cdot k_{3R}\ln\left(
1-\bar{z}\right)  \right\} \nonumber\\
&  =\int d^{2}z\exp\left\{  2k_{1R}\cdot k_{2R}\ln\left\vert z\right\vert
+2k_{2R}\cdot k_{3R}\ln\left\vert 1-z\right\vert \right. \nonumber\\
&  \text{ \ \ \ \ \ \ \ \ \ \ \ \ \ \ \ \ }\left.  +\left(  k_{1L}\cdot
k_{2L}-k_{1R}\cdot k_{2R}\right)  \ln z+\left(  k_{2L}\cdot k_{3L}-k_{2R}\cdot
k_{3R}\right)  \ln\left(  1-z\right)  \right\}  ,
\end{align}
where we have used $\alpha^{\prime}=2$ for closed string propagators%
\begin{align}
\left\langle X\left(  z\right)  X\left(  z^{\prime}\right)  \right\rangle  &
=-\dfrac{\alpha^{\prime}}{2}\ln\left(  z-z^{\prime}\right)  ,\\
\left\langle \tilde{X}\left(  \bar{z}\right)  \tilde{X}\left(  \bar{z}%
^{\prime}\right)  \right\rangle  &  =-\dfrac{\alpha^{\prime}}{2}\ln\left(
\bar{z}-\bar{z}^{\prime}\right)  .
\end{align}
Note that for this simple case, Eq.(\ref{mass}) implies either $m=0$ or $n=0$.
However, we will keep track of the general values of $(m,n)$ here for the
reference of future calculations. By using the formula \cite{Heter}%
\begin{align}
I  &  =\int\frac{d^{2}z}{\pi}\left\vert z\right\vert ^{\alpha}\left\vert
1-z\right\vert ^{\beta}z^{n}\left(  1-z\right)  ^{m}\nonumber\\
&  =\frac{\Gamma\left(  -1-\frac{1}{2}\alpha-\frac{1}{2}\beta\right)
\Gamma\left(  1+n+\frac{1}{2}\alpha\right)  \Gamma\left(  1+m+\frac{1}{2}%
\beta\right)  }{\Gamma\left(  -\frac{1}{2}\alpha\right)  \Gamma\left(
-\frac{1}{2}\beta\right)  \Gamma\left(  2+n+m+\frac{1}{2}\alpha+\frac{1}%
{2}\beta\right)  }, \label{Heter}%
\end{align}
we obtain%
\begin{align}
&  A_{\text{closed}}^{(N_{R}+N_{L}=0)}\left(  s,t,u\right) \nonumber\\
&  =\pi\frac{\Gamma\left(  -1-k_{1R}\cdot k_{2R}-k_{2R}\cdot k_{3R}\right)
\Gamma\left(  1+k_{1L}\cdot k_{2L}\right)  \Gamma\left(  1+k_{2L}\cdot
k_{3L}\right)  }{\Gamma\left(  -k_{1R}\cdot k_{2R}\right)  \Gamma\left(
-k_{2R}\cdot k_{3R}\right)  \Gamma\left(  2+k_{1L}\cdot k_{2L}+k_{2L}\cdot
k_{3L}\right)  }\nonumber\\
&  =\frac{\sin\left(  -\pi k_{1R}\cdot k_{2R}\right)  \sin\left(  -\pi
k_{2R}\cdot k_{3R}\right)  }{\sin\left(  -\pi-\pi k_{1R}\cdot k_{2R}-\pi
k_{2R}\cdot k_{3R}\right)  }\nonumber\\
&  \text{ \ \ \ }\cdot\frac{\Gamma\left(  1+k_{1R}\cdot k_{2R}\right)
\Gamma\left(  1+k_{2R}\cdot k_{3R}\right)  }{\Gamma\left(  2+k_{1R}\cdot
k_{2R}+k_{2R}\cdot k_{3R}\right)  }\frac{\Gamma\left(  1+k_{1L}\cdot
k_{2L}\right)  \Gamma\left(  1+k_{2L}\cdot k_{3L}\right)  }{\Gamma\left(
2+k_{1L}\cdot k_{2L}+k_{2L}\cdot k_{3L}\right)  }\nonumber\\
&  \simeq\frac{\sin\left(  \pi s_{L}/2\right)  \sin\left(  \pi t_{R}/2\right)
}{\sin\left(  \pi u_{L}/2\right)  }\frac{\Gamma\left(  -1-\frac{t_{R}}%
{2}\right)  \Gamma\left(  -1-\frac{u_{R}}{2}\right)  }{\Gamma\left(
2+\frac{s_{R}}{2}\right)  }\frac{\Gamma\left(  -1-\frac{t_{L}}{2}\right)
\Gamma\left(  -1-\frac{u_{L}}{2}\right)  }{\Gamma\left(  2+\frac{s_{L}}%
{2}\right)  }, \label{4-tachyon}%
\end{align}
where we have used $M_{iL,R}^{2}=-2$ for $i=1,2,3,4$. In the above
calculation, we have used the following well known formula for gamma function%
\begin{equation}
\Gamma\left(  x\right)  =\frac{\pi}{\sin\left(  \pi x\right)  \Gamma\left(
1-x\right)  }. \label{gamma}%
\end{equation}
%

\setcounter{equation}{0}
\renewcommand{\theequation}{\arabic{section}.\arabic{equation}}%

\section{High energy behaviors}

\subsection{High energy massive scatterings for general $N_{R}+N_{L}$}

We now proceed to calculate the high energy scattering amplitudes for general
higher mass levels with fixed $N_{R}+N_{L}$. With one compactified coordinate,
the mass spectrum of the second vertex of the amplitude is%
\begin{equation}
M_{2}^{2}=\left(  \dfrac{m_{2}^{2}}{R^{2}}+\dfrac{1}{4}n_{2}^{2}R^{2}\right)
+N_{R}+N_{L}-2.
\end{equation}%
\begin{figure}
[ptb]
\begin{center}
\includegraphics[
height=2.258in,
width=3.5898in
]%
{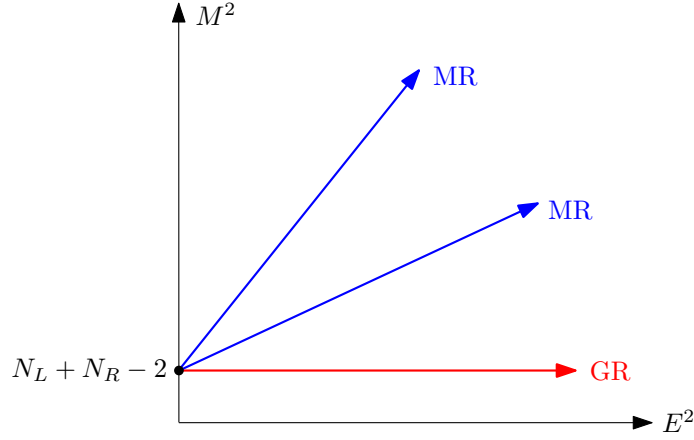}%
\caption{Different regimes of "high energy limit". The high energy regime
defined by $E^{2}\simeq M_{2}^{2}$ $\gg$ $N_{R}+N_{L}$ will be called Mende
regime (MR). The high energy regime defined by $E^{2}\gg M_{2}^{2}$, $E^{2}%
\gg$ $N_{R}+N_{L}$ will be called Gross regime (GR).}%
\label{GR-MR}%
\end{center}
\end{figure}
We now have more mass parameters to define the "high energy limit". So let's
first clear and redefine the concept of "high energy limit" in our following
calculations. We are going to use three quantities $E^{2},M_{2}^{2}$ and
$N_{R}+N_{L}$ to define different regimes of "high energy limit". See FIG.
\ref{GR-MR}. The high energy regime defined by $E^{2}\simeq M_{2}^{2}$ $\gg$
$N_{R}+N_{L}$ will be called Mende regime (MR). The high energy regime defined
by $E^{2}\gg M_{2}^{2}$, $E^{2}\gg$ $N_{R}+N_{L}$ will be called Gross region
(GR). In the high energy limit, the polarizations on the scattering plane for
the second vertex operator are defined to be%

\begin{align}
e^{\mathbf{P}}  &  =\frac{1}{M_{2}}\left(  \sqrt{p^{2}+M_{2}^{2}%
},p,0,0\right)  ,\\
e^{\mathbf{L}}  &  =\frac{1}{M_{2}}\left(  p,\sqrt{p^{2}+M_{2}^{2}%
},0,0\right)  ,\\
e^{\mathbf{T}}  &  =\left(  0,0,1,0\right)
\end{align}
where the fourth component refers to the compactified direction. One can
calculate the following formulas in the high energy limit%
\begin{align}
e^{\mathbf{P}}\cdot k_{1L}  &  =e^{\mathbf{P}}\cdot k_{1R}=-\frac{1}{M_{2}%
}\left(  \sqrt{p^{2}+M_{1}^{2}}\sqrt{p^{2}+M_{2}^{2}}+p^{2}\right) \nonumber\\
&  =-\frac{p^{2}}{M_{2}}\left(  2+\frac{M_{1}^{2}}{2p^{2}}+\frac{M_{2}^{2}%
}{2p^{2}}\right)  +\mathcal{O}(p^{-2}),\label{K1}\\
e^{\mathbf{P}}\cdot k_{3L}  &  =e^{\mathbf{P}}\cdot k_{3R}=\frac{1}{M_{2}%
}\left(  \sqrt{q^{2}+M_{3}^{2}}\sqrt{p^{2}+M_{2}^{2}}-pq\cos\phi\right)
\nonumber\\
&  =\frac{pq}{M_{2}}\left[  1-\cos\phi+\frac{M_{2}^{2}}{2p^{2}}+\frac
{M_{3}^{2}}{2q^{2}}\right]  +\mathcal{O}(p^{-2}),
\end{align}%
\begin{align}
e^{\mathbf{L}}\cdot k_{1L}  &  =e^{\mathbf{L}}\cdot k_{1R}=-\frac{p}{M_{2}%
}\left(  \sqrt{p^{2}+M_{1}^{2}}+\sqrt{p^{2}+M_{2}^{2}}\right) \nonumber\\
&  =-\frac{p^{2}}{M_{2}}\left(  2+\frac{M_{1}^{2}}{2p^{2}}+\frac{M_{2}^{2}%
}{2p^{2}}\right)  +\mathcal{O}(p^{-2}),\\
e^{\mathbf{L}}\cdot k_{3L}  &  =e^{\mathbf{L}}\cdot k_{3R}=\frac{1}{M_{2}%
}\left(  p\sqrt{q^{2}+M_{3}^{2}}-q\sqrt{p^{2}+M_{2}^{2}}\cos\phi\right)
\nonumber\\
&  =\frac{pq}{M_{2}}\left[  1+\frac{M_{3}^{2}}{2q^{2}}-\left(  1+\frac
{M_{2}^{2}}{2p^{2}}\right)  \cos\phi\right]  +\mathcal{O}(p^{-2}), \label{K8}%
\end{align}%
\begin{align}
e^{\mathbf{T}}\cdot k_{1L}  &  =e^{\mathbf{T}}\cdot k_{1R}=0,\\
e^{\mathbf{T}}\cdot k_{3L}  &  =e^{\mathbf{T}}\cdot k_{3R}=-q\sin\phi,
\label{K2}%
\end{align}
which will be useful in the calculations of high energy string scattering amplitudes.

For the noncompactified open string, it was shown that \cite{CHLTY,PRL}, at
each fixed mass level $M_{op}^{2}=2(N-1)$, a four-point function is at the
leading order in high energy (GR) only for states of the following form
\begin{equation}
\left\vert N,2l,q\right\rangle \equiv(\alpha_{-1}^{\mathbf{T}})^{N-2l-2q}%
(\alpha_{-1}^{\mathbf{L}})^{2l}(\alpha_{-2}^{\mathbf{L}})^{q}\left\vert
0,k\right\rangle
\end{equation}
where $N\geqslant2l+2q,l,q\geqslant0$. To avoid the complicated subleading
order calculation due to the $\alpha_{-1}^{\mathbf{L}}$ operator, we will
choose the simple case $l=0$. We made a similar choice when dealing with the
high energy string/D-brane scatterings \cite{Dscatt, Wall, Decay}. There is
still one complication in the case of compactified string due to the possible
choices of $\alpha_{-n}^{25}$ and $\tilde{\alpha}_{-m}^{25}$ in the vertex
operator. However, it can be easily shown that for each fixed mass level with
given quantized and winding momenta $(\frac{m}{R},\frac{1}{2}nR)$, and thus
fixed $N_{R}+N_{L}$ level, vertex operators containing $\alpha_{-n}^{25}$ or
$\tilde{\alpha}_{-m}^{25}$ are subleading order in energy in the high energy
expansion compared to other choices $\alpha_{-1}^{\mathbf{T}}(\tilde{\alpha
}_{-1}^{\mathbf{T}})$ and $\alpha_{-2}^{\mathbf{L}}$ $(\tilde{\alpha}%
_{-2}^{\mathbf{L}})$ on the noncompact directions. In conclusion, in the
calculation of compactified closed string in the GR, we are going to consider
tensor state of the form%
\begin{equation}
\left\vert N_{L,R},q_{L,R}\right\rangle \equiv\left(  \alpha_{-1}^{\mathbf{T}%
}\right)  ^{N_{L}-2q_{L}}\left(  \alpha_{-2}^{\mathbf{L}}\right)  ^{q_{L}%
}\otimes\left(  \tilde{\alpha}_{-1}^{\mathbf{T}}\right)  ^{N_{R}-2q_{R}%
}\left(  \tilde{\alpha}_{-2}^{\mathbf{L}}\right)  ^{q_{R}}\left\vert
0\right\rangle , \label{tensor}%
\end{equation}
at general $N_{R}+N_{L}$ level scattered from three other tachyon states with
$N_{R}+N_{L}=0$.

Note that, in the GR, one can identify $e^{\mathbf{P}}$ with $e^{\mathbf{L}}$
as usual \cite{ChanLee1,ChanLee2}. However, in the MR, one can not identify
$e^{\mathbf{P}}$ with $e^{\mathbf{L}}$. This can be seen from Eq.(\ref{K1}) to
Eq.(\ref{K8}). In the MR, instead of using the tensor vertex in
Eq.(\ref{tensor}), we will use%
\begin{equation}
\left\vert N_{L,R},q_{L,R}\right\rangle \equiv\left(  \alpha_{-1}^{\mathbf{T}%
}\right)  ^{N_{L}-2q_{L}}\left(  \alpha_{-2}^{\mathbf{P}}\right)  ^{q_{L}%
}\otimes\left(  \tilde{\alpha}_{-1}^{\mathbf{T}}\right)  ^{N_{R}-2q_{R}%
}\left(  \tilde{\alpha}_{-2}^{\mathbf{P}}\right)  ^{q_{R}}\left\vert
0\right\rangle , \label{new}%
\end{equation}
as the second vertex operator in the calculation of high energy scattering
amplitudes. Note also that, in the MR, states in Eq.(\ref{new}) may not be the
only states which contribute to the high energy scattering amplitudes as in
the GR. However, we will just choose these states to calculate the scattering
amplitudes in order to compare with the corresponding high energy scattering
amplitudes in the GR.

The high energy scattering amplitudes in the MR can be calculated to be
\begin{align}
A  &  =\varepsilon_{\mathbf{T}^{N_{L}-2q_{L}}\mathbf{P}^{q_{L}},\mathbf{T}%
^{N_{R}-2q_{R}}\mathbf{P}^{q_{R}}}\int d^{2}z_{1}d^{2}z_{2}d^{2}z_{3}%
d^{2}z_{4}\nonumber\\
&  \cdot\left\langle V_{1}\left(  z_{1},\bar{z}_{1}\right)  V_{2}%
^{\mathbf{T}^{N_{L}-2q_{L}}\mathbf{P}^{q_{L}},\mathbf{T}^{N_{R}-2q_{R}%
}\mathbf{P}^{q_{R}}}\left(  z_{2},\bar{z}_{2}\right)  V_{3}\left(  z_{3}%
,\bar{z}_{3}\right)  V_{4}\left(  z_{4},\bar{z}_{4}\right)  \right\rangle
\nonumber\\
&  =\varepsilon_{\mathbf{T}^{N_{L}-2q_{L}}\mathbf{P}^{q_{L}},\mathbf{T}%
^{N_{R}--2q_{R}}\mathbf{P}^{q_{R}}}\int d^{2}z_{1}d^{2}z_{2}d^{2}z_{3}%
d^{2}z_{4}\left\langle e^{ik_{1L}X}\left(  z_{1}\right)  e^{ik_{1R}\tilde{X}%
}\left(  \bar{z}_{1}\right)  \right. \nonumber\\
&  \left.  \cdot\left(  \partial X^{\mathbf{T}}\right)  ^{N_{L}-2q_{L}}\left(
i\partial^{2}X^{\mathbf{P}}\right)  ^{q_{L}}e^{ik_{2L}X}\left(  z_{2}\right)
\left(  \bar{\partial}\tilde{X}^{\mathbf{T}}\right)  ^{N_{R}-2q_{R}}\left(
i\bar{\partial}^{2}\tilde{X}^{\mathbf{P}}\right)  ^{q_{R}}e^{ik_{2R}\tilde{X}%
}\left(  \bar{z}_{2}\right)  \right. \nonumber\\
&  \left.  e^{ik_{3L}X}\left(  z_{3}\right)  e^{ik_{3R}\tilde{X}}\left(
\bar{z}_{3}\right)  e^{ik_{4L}X}\left(  z_{4}\right)  e^{ik_{4R}\tilde{X}%
}\left(  \bar{z}_{4}\right)  \right\rangle \nonumber\\
&  =\int d^{2}z_{1}d^{2}z_{2}d^{2}z_{3}d^{2}z_{4}\cdot\left[  \prod
\limits_{i<j}\left(  z_{i}-z_{j}\right)  ^{k_{iL}\cdot k_{jL}}\left(  \bar
{z}_{i}-\bar{z}_{j}\right)  ^{k_{iR}\cdot k_{jR}}\right] \nonumber\\
&  \cdot\left[  \frac{ie^{\mathbf{T}}\cdot k_{1L}}{z_{1}-z_{2}}+\frac
{ie^{\mathbf{T}}\cdot k_{3L}}{z_{3}-z_{2}}+\frac{ie^{\mathbf{T}}\cdot k_{4L}%
}{z_{4}-z_{2}}\right]  ^{N_{L}-2q_{L}}\cdot\left[  \frac{e^{\mathbf{P}}\cdot
k_{1L}}{\left(  z_{1}-z_{2}\right)  ^{2}}+\frac{e^{\mathbf{P}}\cdot k_{3L}%
}{\left(  z_{3}-z_{2}\right)  ^{2}}+\frac{e^{\mathbf{P}}\cdot k_{4L}}{\left(
z_{4}-z_{2}\right)  ^{2}}\right]  ^{q_{L}}\nonumber\\
&  \cdot\left[  \frac{ie^{\mathbf{T}}\cdot k_{1R}}{\bar{z}_{1}-\bar{z}_{2}%
}+\frac{ie^{\mathbf{T}}\cdot k_{3R}}{\bar{z}_{3}-\bar{z}_{2}}+\frac
{ie^{\mathbf{T}}\cdot k_{4R}}{\bar{z}_{4}-\bar{z}_{2}}\right]  ^{N_{R}-2q_{R}%
}\cdot\left[  \frac{e^{\mathbf{P}}\cdot k_{1R}}{\left(  \bar{z}_{1}-\bar
{z}_{2}\right)  ^{2}}+\frac{e^{\mathbf{P}}\cdot k_{3R}}{\left(  \bar{z}%
_{3}-\bar{z}_{2}\right)  ^{2}}+\frac{e^{\mathbf{P}}\cdot k_{4R}}{\left(
\bar{z}_{4}-\bar{z}_{2}\right)  ^{2}}\right]  ^{q_{R}}.
\end{align}
\ After the standard $SL(2,C)$ gauge fixing, we get%
\begin{align}
A  &  \simeq\left(  -1\right)  ^{k_{1L}\cdot k_{2L}+k_{1R}\cdot k_{2R}%
+k_{1L}\cdot k_{3L}+k_{1R}\cdot k_{3R}+k_{2L}\cdot k_{3L}+k_{2R}\cdot k_{3R}%
}\nonumber\\
&  \cdot\int d^{2}z\cdot z^{k_{1L}\cdot k_{2L}}\cdot\bar{z}^{k_{1R}\cdot
k_{2R}}\cdot\left(  1-z\right)  ^{k_{2L}\cdot k_{3L}}\left(  1-\bar{z}\right)
^{k_{2R}\cdot k_{3R}}\nonumber\\
&  \cdot\left[  \frac{ie^{\mathbf{T}}\cdot k_{1L}}{z}-\frac{ie^{\mathbf{T}%
}\cdot k_{3L}}{1-z}\right]  ^{N_{L}-2q_{L}}\cdot\left[  \frac{ie^{\mathbf{T}%
}\cdot k_{1R}}{\bar{z}}-\frac{ie^{\mathbf{T}}\cdot k_{3R}}{1-\bar{z}}\right]
^{N_{R}-2q_{R}}\nonumber\\
&  \cdot\left[  \frac{e^{\mathbf{P}}\cdot k_{1L}}{z^{2}}+\frac{e^{\mathbf{P}%
}\cdot k_{3L}}{\left(  1-z\right)  ^{2}}\right]  ^{q_{L}}\cdot\left[
\frac{e^{\mathbf{P}}\cdot k_{1R}}{\bar{z}^{2}}+\frac{e^{\mathbf{P}}\cdot
k_{3R}}{\left(  1-\bar{z}\right)  ^{2}}\right]  ^{q_{R}}.
\end{align}
By using Eqs.(\ref{K1}) to (\ref{K2}), the amplitude can be written as%

\begin{align*}
A  &  \sim\left(  -1\right)  ^{n+q+q^{\prime}+k_{1L}\cdot k_{2L}+k_{1R}\cdot
k_{2R}+k_{1L}\cdot k_{3L}+k_{1R}\cdot k_{3R}+k_{2L}\cdot k_{3L}+k_{2R}\cdot
k_{3R}}\left(  q\sin\phi\right)  ^{N_{L}+N_{R}-2q_{L}-2q_{R}}\\
&  \cdot\int d^{2}z\cdot z^{k_{1L}\cdot k_{2L}}\cdot\bar{z}^{k_{1R}\cdot
k_{2R}}\cdot\left(  1-z\right)  ^{k_{2L}\cdot k_{3L}}\left(  1-\bar{z}\right)
^{k_{2R}\cdot k_{3R}}\cdot\left[  \frac{1}{1-z}\right]  ^{N_{L}-2q_{L}}\left[
\frac{1}{1-\bar{z}}\right]  ^{N_{R}-2q_{R}}\\
&  \cdot\left[  \frac{-\frac{1}{M_{2}}\left(  \sqrt{p^{2}+M_{1}^{2}}%
\sqrt{p^{2}+M_{2}^{2}}+p^{2}\right)  }{z^{2}}+\frac{\frac{1}{M_{2}}\left(
\sqrt{q^{2}+M_{3}^{2}}\sqrt{p^{2}+M_{2}^{2}}-pq\cos\phi\right)  }{\left(
1-z\right)  ^{2}}\right]  ^{q_{L}}\\
&  \cdot\left[  \frac{-\frac{1}{M_{2}}\left(  \sqrt{p^{2}+M_{1}^{2}}%
\sqrt{p^{2}+M_{2}^{2}}+p^{2}\right)  }{\bar{z}^{2}}+\frac{\frac{1}{M_{2}%
}\left(  \sqrt{q^{2}+M_{3}^{2}}\sqrt{p^{2}+M_{2}^{2}}-pq\cos\phi\right)
}{\left(  1-\bar{z}\right)  ^{2}}\right]  ^{q_{R}}.
\end{align*}

\begin{align}
&  =\left(  -1\right)  ^{k_{1L}\cdot k_{2L}+k_{1R}\cdot k_{2R}+k_{1L}\cdot
k_{3L}+k_{1R}\cdot k_{3R}+k_{2L}\cdot k_{3L}+k_{2R}\cdot k_{3R}}\left(
q\sin\phi\right)  ^{N_{L}+N_{R}}\left(  \frac{1}{M_{2}q^{2}\sin^{2}\phi
}\right)  ^{q_{L}+q_{R}}\nonumber\\
&  \cdot\int d^{2}z\cdot z^{k_{1L}\cdot k_{2L}}\cdot\bar{z}^{k_{1R}\cdot
k_{2R}}\cdot\left(  1-z\right)  ^{k_{2L}\cdot k_{3L}}\left(  1-\bar{z}\right)
^{k_{2R}\cdot k_{3R}}\cdot\left[  \frac{1}{1-z}\right]  ^{N_{L}-2q_{L}}\left[
\frac{1}{1-\bar{z}}\right]  ^{N_{R}-2q_{R}}\nonumber\\
&  \cdot\sum_{i=0}^{q}\sum_{j=0}^{q^{\prime}}\binom{q}{i}\binom{q^{\prime}}%
{j}\left(  \frac{\sqrt{p^{2}+M_{1}^{2}}\sqrt{p^{2}+M_{2}^{2}}+p^{2}}{z^{2}%
}\right)  ^{i}\left(  \frac{\sqrt{p^{2}+M_{1}^{2}}\sqrt{p^{2}+M_{2}^{2}}%
+p^{2}}{\bar{z}^{2}}\right)  ^{j}\nonumber\\
&  =\left(  -1\right)  ^{k_{1L}\cdot k_{2L}+k_{1R}\cdot k_{2R}+k_{1L}\cdot
k_{3L}+k_{1R}\cdot k_{3R}+k_{2L}\cdot k_{3L}+k_{2R}\cdot k_{3R}}\left(
q\sin\phi\right)  ^{N_{L}+N_{R}}\nonumber\\
&  \cdot\left(  -\frac{\sqrt{q^{2}+M_{3}^{2}}\sqrt{p^{2}+M_{2}^{2}}-pq\cos
\phi}{M_{2}q^{2}\sin^{2}\phi}\right)  ^{q_{L}+q_{R}}\nonumber\\
&  \cdot\sum_{i=0}^{q_{L}}\sum_{j=0}^{q_{R}}\binom{q_{L}}{i}\binom{q_{R}}%
{j}\left(  \frac{\sqrt{p^{2}+M_{1}^{2}}\sqrt{p^{2}+M_{2}^{2}}+p^{2}}%
{-\sqrt{q^{2}+M_{3}^{2}}\sqrt{p^{2}+M_{2}^{2}}+pq\cos\phi}\right)
^{i+j}\nonumber\\
&  \cdot\frac{\sin\left[  -\pi\left(  k_{1R}\cdot k_{2R}-2j\right)  \right]
\sin\left[  -\pi\left(  k_{2R}\cdot k_{3R}-N_{R}+2j\right)  \right]  }%
{\sin\left[  -\pi\left(  1+k_{1R}\cdot k_{2R}+k_{2R}\cdot k_{3R}-N_{R}\right)
\right]  }\nonumber\\
&  \cdot\frac{\Gamma\left(  1+k_{1R}\cdot k_{2R}-2j\right)  \Gamma\left(
1+k_{2R}\cdot k_{3R}-N_{R}+2j\right)  }{\Gamma\left(  2+k_{1R}\cdot
k_{2R}+k_{2R}\cdot k_{3R}-N_{R}\right)  }\nonumber\\
&  \cdot\frac{\Gamma\left(  1+k_{1L}\cdot k_{2L}-2i\right)  \Gamma\left(
1+k_{2L}\cdot k_{3L}+2i-N_{L}\right)  }{\Gamma\left(  2+k_{1L}\cdot
k_{2L}+k_{2L}\cdot k_{3L}-N_{L}\right)  },
\end{align}
where, as in the calculation of section II for the GR, we have used
Eq.(\ref{Heter}) to do the integration. It is easy to do the following
approximations for the gamma functions%
\begin{align}
&  A\simeq\left(  -1\right)  ^{k_{1L}\cdot k_{2L}+k_{1R}\cdot k_{2R}%
+k_{1L}\cdot k_{3L}+k_{1R}\cdot k_{3R}+k_{2L}\cdot k_{3L}+k_{2R}\cdot k_{3R}%
}\left(  q\sin\phi\right)  ^{N_{L}+N_{R}}\nonumber\\
&  \cdot\left(  -\frac{\sqrt{q^{2}+M_{3}^{2}}\sqrt{p^{2}+M_{2}^{2}}-pq\cos
\phi}{M_{2}q^{2}\sin^{2}\phi}\right)  ^{q_{L}+q_{R}}\nonumber\\
&  \cdot\sum_{i=0}^{q_{L}}\sum_{j=0}^{q_{R}}\binom{q_{L}}{i}\binom{q_{R}}%
{j}\left(  \frac{\sqrt{p^{2}+M_{1}^{2}}\sqrt{p^{2}+M_{2}^{2}}+p^{2}}%
{-\sqrt{q^{2}+M_{3}^{2}}\sqrt{p^{2}+M_{2}^{2}}+pq\cos\phi}\right)
^{i+j}\nonumber\\
&  \cdot\frac{\sin\left[  -\pi k_{1R}\cdot k_{2R}\right]  \sin\left[  -\pi
k_{2R}\cdot k_{3R}\right]  }{\sin\left[  -\pi\left(  1+k_{1R}\cdot
k_{2R}+k_{2R}\cdot k_{3R}\right)  \right]  }\nonumber\\
&  \cdot\frac{\Gamma\left(  1+k_{1R}\cdot k_{2R}\right)  \Gamma\left(
1+k_{2R}\cdot k_{3R}\right)  \Gamma\left(  1+k_{1L}\cdot k_{2L}\right)
\Gamma\left(  1+k_{2L}\cdot k_{3L}\right)  }{\Gamma\left(  2+k_{1R}\cdot
k_{2R}+k_{2R}\cdot k_{3R}\right)  \Gamma\left(  2+k_{1L}\cdot k_{2L}%
+k_{2L}\cdot k_{3L}\right)  }\nonumber\\
&  \cdot\dfrac{\left(  k_{1R}\cdot k_{2R}\right)  ^{-2j}\left(  k_{2R}\cdot
k_{3R}\right)  ^{-N_{R}+2j}}{\left(  k_{1R}\cdot k_{2R}+k_{2R}\cdot
k_{3R}\right)  ^{-N_{R}}}\dfrac{\left(  k_{1L}\cdot k_{2L}\right)
^{-2i}\left(  k_{2L}\cdot k_{3L}\right)  ^{-N_{L}+2i}}{\left(  k_{1L}\cdot
k_{2L}+k_{2L}\cdot k_{3L}\right)  ^{-N_{L}}}.
\end{align}
One can now do the double summation\ and drop out the $M_{iL,R}$ terms to get%
\begin{align}
A  &  \simeq\left(  -\dfrac{q\sin\phi\left(  s_{L}+t_{L}\right)  }{t_{L}%
}\right)  ^{N_{L}}\left(  -\dfrac{q\sin\phi\left(  s_{R}+t_{R}\right)  }%
{t_{R}}\right)  ^{N_{R}}\left(  \frac{1}{2M_{2}q^{2}\sin^{2}\phi}\right)
^{q_{L}+q_{R}}\nonumber\\
&  \cdot\left(  \left(  t_{R}-2\vec{K}_{2R}\cdot\vec{K}_{3R}\right)
+\dfrac{t_{R}^{2}\left(  s_{R}-2\vec{K}_{1R}\cdot\vec{K}_{2R}\right)  }%
{s_{R}^{2}}\right)  ^{q_{R}}\nonumber\\
&  \cdot\left(  \left(  t_{L}-2\vec{K}_{2L}\cdot\vec{K}_{3L}\right)
+\dfrac{t_{L}^{2}\left(  s_{L}-2\vec{K}_{1L}\cdot\vec{K}_{2L}\right)  }%
{s_{L}^{2}}\right)  ^{q_{L}}\nonumber\\
&  \cdot\frac{\sin\left(  \pi s_{L}/2\right)  \sin\left(  \pi t_{R}/2\right)
}{\sin\left(  \pi u_{L}/2\right)  }B\left(  -1-\dfrac{t_{R}}{2},-1-\dfrac
{u_{R}}{2}\right)  B\left(  -1-\dfrac{t_{L}}{2},-1-\dfrac{u_{L}}{2}\right)  .
\label{amplitude}%
\end{align}
Eq.(\ref{amplitude}) is valid for $E^{2}\gg N_{R}+N_{L},$ $M_{2}^{2}\gg
N_{R}+N_{L}.$

\subsection{The infinite linear relations in the GR}

For the special case of GR with $E^{2}\gg M_{2}^{2}$, one can identify $q$
with $p$, and the amplitude in Eq.(\ref{amplitude}) further reduces to%
\begin{align}
\lim_{E^{2}\gg M_{2}^{2}}A  &  =\left(  \frac{2p\cos^{3}\frac{\phi}{2}}%
{\sin\frac{\phi}{2}}\right)  ^{N_{L}+N_{R}}\left(  -\frac{1}{2M_{2}}\right)
^{q_{L}+q_{R}}\frac{\sin\left(  \pi s_{L}/2\right)  \sin\left(  \pi
t_{R}/2\right)  }{\sin\left(  \pi u_{L}/2\right)  }\nonumber\\
&  \cdot B\left(  -1-\dfrac{t_{R}}{2},-1-\dfrac{u_{R}}{2}\right)  B\left(
-1-\dfrac{t_{L}}{2},-1-\dfrac{u_{L}}{2}\right)  . \label{beta}%
\end{align}
It is crucial to note that the high energy limit of the beta function with
$s+t+u=2n-8$ is \cite{ChanLee1}%
\begin{align}
B\left(  -1-\dfrac{t}{2},-1-\dfrac{u}{2}\right)   &  =\frac{\Gamma(-\frac
{t}{2}-1)\Gamma(-\frac{u}{2}-1)}{\Gamma(\frac{s}{2}+2)}\nonumber\\
&  \simeq E^{-1-2n}\left(  \sin\frac{\phi}{2}\right)  ^{-3}\left(  \cos
\frac{\phi}{2}\right)  ^{5-2n}\nonumber\\
&  \cdot\exp\left(  -\frac{t\ln t+u\ln u-(t+u)\ln(t+u)}{2}\right)
\end{align}
where we have calculated the approximation up to the next leading order in
energy $E$. Note that the appearance of the prepower factors in front of the
exponential fall-off factor. For our purpose here, with Eq.(\ref{sum}), we
have%
\begin{equation}
s_{L,R}+t_{L,R}+u_{L,R}=\sum_{i}M_{iL,R}^{2}=2N_{L,R}-8,
\end{equation}
and the high energy limit of the beta functions in Eq.(\ref{beta}) can be
further calculated to be%
\begin{align}
&  B\left(  -1-\dfrac{t_{R}}{2},-1-\dfrac{u_{R}}{2}\right)  B\left(
-1-\dfrac{t_{L}}{2},-1-\dfrac{u_{L}}{2}\right) \nonumber\\
&  \simeq E^{-1-2(N_{L}+N_{R})}\left(  \sin\frac{\phi}{2}\right)  ^{-3}\left(
\cos\frac{\phi}{2}\right)  ^{5-2(N_{L}+N_{R})}\nonumber\\
&  \cdot\exp\left(  -\frac{t_{L}\ln t_{L}+u_{L}\ln u_{L}-(t_{L}+u_{L}%
)\ln(t_{L}+u_{L})}{2}\right) \nonumber\\
&  \cdot\exp\left(  -\frac{t_{R}\ln t_{R}+u_{R}\ln u_{R}-(t_{R}+u_{R}%
)\ln(t_{R}+u_{R})}{2}\right) \nonumber\\
&  \simeq E^{-1-2(N_{L}+N_{R})}\left(  \sin\frac{\phi}{2}\right)  ^{-3}\left(
\cos\frac{\phi}{2}\right)  ^{5-2(N_{L}+N_{R})}\exp\left(  -\frac{t\ln t+u\ln
u-(t+u)\ln(t+u)}{4}\right)  , \label{final}%
\end{align}
where we have implicitly used the relation $\alpha_{\text{closed}}^{\prime
}=4\alpha_{\text{open}}^{\prime}=2.$ By combining Eq.(\ref{beta}) and
Eq.(\ref{final}), we end up with%
\begin{align}
\lim_{E^{2}\gg M_{2}^{2}}A  &  \simeq\left(  -\frac{2\cot\frac{\phi}{2}}%
{E}\right)  ^{N_{L}+N_{R}}\left(  -\frac{1}{2M_{2}}\right)  ^{q_{L}+q_{R}%
}E^{-1}\left(  \sin\frac{\phi}{2}\right)  ^{-3}\left(  \cos\frac{\phi}%
{2}\right)  ^{5}\nonumber\\
&  \cdot\frac{\sin\left(  \pi s_{L}/2\right)  \sin\left(  \pi t_{R}/2\right)
}{\sin\left(  \pi u_{L}/2\right)  }\exp\left(  -\frac{t\ln t+u\ln
u-(t+u)\ln(t+u)}{4}\right)  . \label{linear}%
\end{align}
We see that there is a $\left(  \frac{m}{R},\frac{1}{2}nR\right)  $ dependence
in the $\frac{\sin\left(  \pi s_{L}/2\right)  \sin\left(  \pi t_{R}/2\right)
}{\sin\left(  \pi u_{L}/2\right)  }$ factor in our final result. This is
physically consistent as one expects a $\left(  \frac{m}{R},\frac{1}%
{2}nR\right)  $ dependent Regge-pole and zero structures in the high energy
string scattering amplitudes. In conclusion, in the GR, for each fixed mass
level with given quantized and winding momenta $\left(  \frac{m}{R},\frac
{1}{2}nR\right)  $ (thus fixed $N_{L}$ and $N_{R}$ by Eq.(\ref{mass})), we
have obtained infinite linear relations among high energy scattering
amplitudes of different string states with various $(q_{L},q_{R})$. Note also
that this result reproduces the correct ratios $\left(  -\frac{1}{2M_{2}%
}\right)  ^{q_{L}+q_{R}}$ obtained in the previous works \cite{Dscatt, Wall,
Decay}. However, the mass parameter $M_{2}$ here depends on $\left(  \frac
{m}{R},\frac{1}{2}nR\right)  $. It is also interesting to see that, if not for
the $\left(  \frac{m}{R},\frac{1}{2}nR\right)  $ dependence in the $\frac
{\sin\left(  \pi s_{L}/2\right)  \sin\left(  \pi t_{R}/2\right)  }{\sin\left(
\pi u_{L}/2\right)  }$ factor in the high energy scattering amplitudes in the
GR, we would have had a linear relation among scattering amplitudes of
different string states in different mass levels with fixed $\left(
N_{R}+N_{L}\right)  $.

Presumably, the infinite linear relations obtained above can be reproduced by
using the method of high energy ZNS, or high energy Ward identities, adopted
in the previous works \cite{ChanLee1,ChanLee2,
CHL,CHLTY,PRL,paperB,susy,Closed}. The existence of Soliton ZNS at arbitrary
mass levels was constructed in \cite{Lee}. A closer look in this direction
seems worthwhile. In the paper, however, we are more interested in
understanding the power-law behavior of the high energy string scattering
amplitudes and breakdown of the infinite linear relations as we will discuss
in the next section.

\subsection{Power-law and breakdown of the infinite linear relations in the
MR}

In this section we discuss the power-law behavior of high energy string
scattering amplitudes in a compact space. We will see that, in the MR, the
infinite linear relations derived in section IIIB \textit{break down} and,
simultaneously, the UV exponential fall-off behavior of high energy string
scattering amplitudes \textit{enhances} to power-law behavior. The power-law
behavior of high energy string scatterings in a compact space was first
suggested by Mende \cite{Mende}. Here we give a mathematically more concrete
description. It is easy to see that the "power law" condition, i.e. Eq.(3.7)
in Mende's paper \cite{Mende},%
\begin{equation}
k_{1L}\cdot k_{2L}+k_{1R}\cdot k_{2R}=\text{constant,}
\label{mandy's condition}%
\end{equation}
turns out to be%
\begin{align}
&  -\left(  k_{1L}\cdot k_{2L}+k_{1R}\cdot k_{2R}\right) \nonumber\\
&  =\sqrt{p^{2}+M_{1}^{2}}\cdot\sqrt{p^{2}+M_{2}^{2}}+p^{2}+\left(  \vec
{K}_{1L}\cdot\vec{K}_{2L}+\vec{K}_{1R}\cdot\vec{K}_{2R}\right) \nonumber\\
&  =\sqrt{p^{2}+M_{1}^{2}}\cdot\sqrt{p^{2}+M_{2}^{2}}+p^{2}+2\left(  \vec
{K}_{1}\cdot\vec{K}_{2}+\vec{L}_{1}\cdot\vec{L}_{2}\right) \nonumber\\
&  =\text{constant}.
\end{align}
As $p\rightarrow\infty$, due to the existence of winding modes in the
compactified closed string, it is possible to choose $\left(  \vec{K}_{1}%
,\vec{K}_{2};\vec{L}_{1},\vec{L}_{2}\right)  $\ such that%
\begin{equation}
\vec{K}_{1}\cdot\vec{K}_{2}+\vec{L}_{1}\cdot\vec{L}_{2}<0,
\end{equation}
and let $\left(  \vec{K}_{1}\cdot\vec{K}_{2}+\vec{L}_{1}\cdot\vec{L}%
_{2}\right)  \rightarrow-\infty$ to make
\begin{align}
k_{1L}\cdot k_{2L}+k_{1R}\cdot k_{2R}\simeq &  \text{ constant}\\
\Rightarrow s_{L}+s_{R}\simeq &  \text{ constant.}%
\end{align}
In our calculation, this condition implies the beta functions in
Eq.(\ref{amplitude}) reduce to%
\begin{align}
&  B\left(  -1-\dfrac{t_{R}}{2},-1-\dfrac{u_{R}}{2}\right)  B\left(
-1-\dfrac{t_{L}}{2},-1-\dfrac{u_{L}}{2}\right) \nonumber\\
&  =\frac{\Gamma(-\frac{t_{R}}{2}-1)\Gamma(-\frac{u_{R}}{2}-1)}{\Gamma
(\frac{s_{R}}{2}+2)}\frac{\Gamma(-\frac{t_{L}}{2}-1)\Gamma(-\frac{u_{L}}%
{2}-1)}{\Gamma(\frac{s_{L}}{2}+2)}\nonumber\\
&  =\frac{\sin\left(  \pi s_{R}/2\right)  \Gamma(-\frac{t_{R}}{2}%
-1)\Gamma(-\frac{u_{R}}{2}-1)\Gamma(-\frac{t_{L}}{2}-1)\Gamma(-\frac{u_{L}}%
{2}-1)}{\pi\frac{s_{R}}{2}\left(  1+\frac{s_{R}}{2}\right)  \left(
-1+\frac{s_{R}}{2}\right)  },
\end{align}
which behaves as \textit{power-law} in the high energy limit! On the other
hand, it is obvious that the $(q_{L},q_{R})$ dependent power factors of the
amplitude in Eq.(\ref{amplitude})%
\begin{align}
A_{q_{L},q_{R}}  &  \simeq\left(  \frac{1}{2M_{2}q^{2}\sin^{2}\phi}\right)
^{q_{L}+q_{R}}\nonumber\\
&  \cdot\left(  \left(  t_{R}-2\vec{K}_{2R}\cdot\vec{K}_{3R}\right)
+\dfrac{t_{R}^{2}\left(  s_{R}-2\vec{K}_{1R}\cdot\vec{K}_{2R}\right)  }%
{s_{R}^{2}}\right)  ^{q_{R}}\nonumber\\
&  \cdot\left(  \left(  t_{L}-2\vec{K}_{2L}\cdot\vec{K}_{3L}\right)
+\dfrac{t_{L}^{2}\left(  s_{L}-2\vec{K}_{1L}\cdot\vec{K}_{2L}\right)  }%
{s_{L}^{2}}\right)  ^{q_{L}}.
\end{align}
show \textit{no} linear relations in the MR. This is very different from the
case of high energy scattering amplitude in Eq.(\ref{linear}) in the GR ,
which shows nice linear relations%
\begin{equation}
A_{q_{L},q_{R}}\simeq\left(  -\frac{1}{2M_{2}}\right)  ^{q_{L}+q_{R}}.
\end{equation}
Note that the mechanism to break the linear relations and the mechanism to
enhance the amplitude to power-law are all due to $E\simeq M_{2}$ in the MR.
In our notation, Eq.(\ref{mandy's condition}) is equivalent to the following
condition%
\begin{equation}
\lim_{p\rightarrow\infty}\frac{\sqrt{p^{2}+M_{1}^{2}}\cdot\sqrt{p^{2}%
+M_{2}^{2}}+p^{2}}{\vec{K}_{1}\cdot\vec{K}_{2}+\vec{L}_{1}\cdot\vec{L}_{2}%
}\sim\frac{E^{2}}{\left(  \dfrac{m_{1}m_{2}}{R^{2}}+\dfrac{1}{4}n_{1}%
n_{2}R^{2}\right)  }\sim-\text{ }\mathcal{O}(1). \label{condition}%
\end{equation}
For our purpose here, as we will see soon, it is good enough to choose only
one compactified coordinate to realize Eq.(\ref{condition}). First of all, in
addition to Eq.(\ref{kk}) and Eq.(\ref{wind}), Eq.(\ref{mass}) implies%
\begin{equation}
m_{i}n_{i}=0,i=1,2,3,4\text{ (no sum on }i\text{).}%
\end{equation}
This is because three of the four vertex are tachyons. Also, since we are
going to take $n_{2}$ to infinity with fixed $N_{R}+N_{L}$ in order to satisfy
Eq.(\ref{condition}), we are forced to take $m_{2}=0$. In sum, we can take,
say, $m_{i}=0$ for $i=1,2,3,4,$ and $n_{1}=-n_{2}=-n,n_{3}=-2n,n_{4}=0,$ and
then let $n\rightarrow\infty$ to realize Eq.(\ref{condition}). Note that it is
crucial to choose different sign for $n_{1}$ and $n_{2}$ in order to achieve
the minus sign in Eq.(\ref{condition}). We stress that there are other choices
to realize the condition. One notes that all choices implies%
\begin{equation}
N_{R}=N_{L}.
\end{equation}
It is obvious that one can also compactify more than one coordinate to realize
the Mende condition. We conclude that the high energy scatterings of the
"highly winding string states" of the compactified closed string in the MR
behave as the unusual UV power-law, and the usual linear relations among
scattering amplitudes break down due to the unusual power-law behavior.

\section{Conclusion}

In this paper we calculate high energy scattering amplitudes of closed bosonic
string compactified on torus. We define two regimes of high energy limit, the
Gross regime (GR) and the Mende regime (MR). In the GR, for each fixed mass
level with given quantized and winding momenta $(\frac{m}{R},\frac{1}{2}nR)$,
we obtain infinite linear relations among high energy scattering amplitudes of
different string states. In the MR, we discover that linear relations with
$N_{R}=N_{L}$ break down and, simultaneously, the amplitudes enhance to
power-law behavior instead of the usual exponential fall-off behavior at high
energies. The result of this work gives a concrete example to justify the
coexistence of the linear relations and the softer exponential fall-off
behavior of high energy string scattering amplitudes as we pointed out
previously \cite{Wall,Decay}. It is also reminiscent of our previous work on
the power-law behavior of high energy string/domain-wall scatterings
\cite{Wall}. However, in the case of string/domain-wall scatterings, the high
energy scattering amplitudes behaves as power-law for the whole UV kinematic
regime, and one can not see the transition from UV power-law behavior to the
UV exponential fall-off behavior, neither can one see the transition from
nonlinear relations to the linear relations among high energy scattering
amplitudes of different string states. On the contrary, for the high energy
scatterings of the compactified closed string considered in this paper, one
gets more kinematic variables, namely $(\frac{m}{R},\frac{1}{2}nR),$ to cover
both GR and MR, so that one can see this interesting transition. Through the
observation of this transition, one further confirms that the infinite linear
relations obtained in \cite{ChanLee1,ChanLee2,
CHL,CHLTY,PRL,paperB,susy,Closed} are responsible for the UV softer string
scattering amplitudes than the field theory scatterings.

\begin{acknowledgments}
We would like to thank the hospitality of University of Tokyo at Komaba, where
this work is finalized. We are indebted to Chuan-Tsung Chan, Koji Hashimoto,
Yoichi Kazama, Akitsugs Miwa, Yu Nakayama, Naoto Yokoi and Tamiaki Yoneya for
many of their enlightening discussions. This work is supported in part by the
National Science Council, 50 billions project of Ministry of Educaton and
National Center for Theoretical Science, Taiwan, R.O.C.
\end{acknowledgments}

\end{document}